\documentstyle[psfig,times]{mn}
\newread\epsffilein    
\newif\ifepsffileok    
\newif\ifepsfbbfound   
\newif\ifepsfverbose   
\newdimen\epsfxsize    
\newdimen\epsfysize    
\newdimen\epsftsize    
\newdimen\epsfrsize    
\newdimen\epsftmp      
\newdimen\pspoints     
\def\epsfbox#1{\global\def\epsfllx{72}\global\def\epsflly{72}%
   \global\def\epsfurx{540}\global\def\epsfury{720}%
   \def\lbracket{[}\def\testit{#1}\ifx\testit\lbracket
   \let\next=\epsfgetlitbb\else\let\next=\epsfnormal\fi\next{#1}}%
\def\epsfgetlitbb#1#2 #3 #4 #5]#6{\epsfgrab #2 #3 #4 #5 .\\%
   \epsfsetgraph{#6}}%
\def\epsfnormal#1{\epsfgetbb{#1}\epsfsetgraph{#1}}%
\def\epsfgetbb#1{%
\openin\epsffilein=#1
\ifeof\epsffilein\errmessage{I couldn't open #1, will ignore it}\else
   {\epsffileoktrue \chardef\other=12
    \def\do##1{\catcode`##1=\other}\dospecials \catcode`\ =10
    \loop
       \read\epsffilein to \epsffileline
       \ifeof\epsffilein\epsffileokfalse\else
          \expandafter\epsfaux\epsffileline:. \\%
       \fi
   \ifepsffileok\repeat
   \ifepsfbbfound\else
    \ifepsfverbose\message{No bounding box comment in #1; using defaults}\fi\fi
   }\closein\epsffilein\fi}%
\def\epsfsetgraph#1{%
   \epsfrsize=\epsfury\pspoints
   \advance\epsfrsize by-\epsflly\pspoints
   \epsftsize=\epsfurx\pspoints
   \advance\epsftsize by-\epsfllx\pspoints
   \epsfxsize\epsfsize\epsftsize\epsfrsize
   \ifnum\epsfxsize=0 \ifnum\epsfysize=0
      \epsfxsize=\epsftsize \epsfysize=\epsfrsize
     \else\epsftmp=\epsftsize \divide\epsftmp\epsfrsize
       \epsfxsize=\epsfysize \multiply\epsfxsize\epsftmp
       \multiply\epsftmp\epsfrsize \advance\epsftsize-\epsftmp
       \epsftmp=\epsfysize
       \loop \advance\epsftsize\epsftsize \divide\epsftmp 2
       \ifnum\epsftmp>0
          \ifnum\epsftsize<\epsfrsize\else
             \advance\epsftsize-\epsfrsize \advance\epsfxsize\epsftmp \fi
       \repeat
     \fi
   \else\epsftmp=\epsfrsize \divide\epsftmp\epsftsize
     \epsfysize=\epsfxsize \multiply\epsfysize\epsftmp   
     \multiply\epsftmp\epsftsize \advance\epsfrsize-\epsftmp
     \epsftmp=\epsfxsize
     \loop \advance\epsfrsize\epsfrsize \divide\epsftmp 2
     \ifnum\epsftmp>0
        \ifnum\epsfrsize<\epsftsize\else
           \advance\epsfrsize-\epsftsize \advance\epsfysize\epsftmp \fi
     \repeat     
   \fi
   \ifepsfverbose\message{#1: width=\the\epsfxsize, height=\the\epsfysize}\fi
   \epsftmp=10\epsfxsize \divide\epsftmp\pspoints
   \newcount\figskipcount
      \message{#1 \the\epsfysize  }
   \vbox to\epsfysize{\vfil\hbox to\epsfxsize{%
      \includegraphics{#1}%
      \hfil}}%
\epsfxsize=0pt\epsfysize=0pt}%

{\catcode`\%=12 \global\let\epsfpercent=
\long\def\epsfaux#1#2:#3\\{\ifx#1\epsfpercent
   \def\testit{#2}\ifx\testit\epsfbblit
      \epsfgrab #3 . . . \\%
      \epsffileokfalse
      \global\epsfbbfoundtrue
   \fi\else\ifx#1\par\else\epsffileokfalse\fi\fi}%
\def\epsfgrab #1 #2 #3 #4 #5\\{%
   \global\def\epsfllx{#1}\ifx\epsfllx\empty
      \epsfgrab #2 #3 #4 #5 .\\\else
   \global\def\epsflly{#2}%
   \global\def\epsfurx{#3}\global\def\epsfury{#4}\fi}%
\def\epsfsize#1#2{\epsfxsize}
\begin{document}
\title[The coincidence and angular clustering of Chandra and SCUBA sources]
{The coincidence and angular clustering of Chandra and SCUBA sources}
\author[O. Almaini et al.]
{
O.~Almaini$^{1}$, 
S.E.~Scott$^{1}$, 
J.S.~Dunlop$^{1}$, 
J.C.~Manners$^{1}$, 
C.J.~Willott$^{2}$, 
A.~Lawrence$^{1}$,\cr
R.J.~Ivison$^{3}$,
O.~Johnson$^{1}$,
A.W.~Blain$^{4}$,
J.A.~Peacock$^{1}$,
S.J.~Oliver$^{5}$,
M.J.~Fox$^{6}$,\cr
R.G.~Mann$^{1}$,
I.~P\'erez-Fournon$^{7}$,
E.~Gonz\'alez-Solares$^{7}$,
M.~Rowan-Robinson$^{6}$,\cr
S.~Serjeant$^{8}$,
F.~Cabrera-Guerra$^{7}$,
D.H.~Hughes$^{9}$
\\
$^1$Institute for Astronomy, 
University of Edinburgh, Royal Observatory, Blackford
Hill, Edinburgh EH9 3HJ\\
$^{2}$ Astrophysics, Department of Physics, Keble Rd, Oxford OX1 3RH\\
$^{3}$ UK Astronomy Technology Centre, Royal Observatory, Blackford Hill, Edinburgh  EH9 3HJ\\
$^{4}$ Institute of Astronomy, Madingley Road, Cambridge  CB3 0HA\\
$^{5}$ Astronomy Centre, CPES, University of Sussex, Falmer, Brighton  
BN1 9QJ\\
$^{6}$ Astrophysics Group, Blackett Laboratory, Imperial College,
Prince Consort Rd.,London SW7 2BW\\
$^{7}$ Instituto de Astrofisica de Canarias, 38200 La Laguna,
Tenerife, Spain\\
$^{8}$ Unit for Space Sciences and Astrophysics, School of Physical
Sciences, University of Kent, Canterbury  CT2 7NZ\\
$^{9}$ Instituto Nacional de Astrofisica, Optica y Electronica (INAOE),
Apartado Postal 51 y 216, 72000 Puebla, Mexico}

\date{MNRAS in press}
\maketitle

\begin{abstract}
We explore the relationship between the hard X-ray and sub-mm
populations using deep Chandra observations of a large, contiguous
SCUBA survey. In agreement with other recent findings, we confirm that
the direct overlap is small. Of the $17$ sub-mm sources detected in
this field at $850~\mu$m, only one is coincident with a Chandra
source. The resulting limits imply that the majority of SCUBA sources
are not powered by AGN, unless the central engine is obscured by
Compton-thick material with a low ($<1$ per cent) scattered component.
Furthermore, since Chandra detects only $\sim 5$ per cent of SCUBA
sources, the typical obscuration would need to be almost isotropic.
The X-ray upper limits are so strong that in most cases we can also
rule out a starburst SED at low redshift, suggesting that the majority
of SCUBA sources lie at $z>1$ even if they are purely starburst
galaxies. Despite the low detection rate, we find evidence for strong
angular clustering between the X-ray and sub-mm populations.  The
implication is that AGN and SCUBA sources trace the same large-scale
structure but do not generally coincide. If bright sub-mm sources
represent massive elliptical galaxies in formation, we suggest that
(for a given galaxy) the major episode of star-formation must be
distinct from the period of observable quasar activity.
\end{abstract}

\begin{keywords} galaxies: active\ -- quasars: general \-- X-rays: general \-- X-rays: galaxies \-- diffuse radiation
\ -- galaxies: evolution \-- galaxies: formation \-- galaxies:
starburst
\end{keywords}

\section{Introduction}

Our understanding of the high-redshift Universe changed dramatically
with the advent of the SCUBA array at the James Clerk Maxwell
Telescope.  A number of groups have since announced the results from
deep submillimetre surveys, all of which come to broadly the same
conclusion.  It appears that a significant (perhaps dominant) fraction
of the star-formation in the high-redshift Universe ($z>2$) took place
in highly dust-enshrouded galaxies (Smail et al. 1997, Hughes et al.
1998, Barger et al. 1998, Eales et al. 1999).  These exceptionally
luminous systems are likely to be the analogues of the Ultra-Luminous
Infrared Galaxies (ULIRGs) seen locally.  The major difference,
however, is their space density. At high redshift ULIRG-like systems
dominate the cosmic energy budget, while locally they are rare and
unusual events. The discovery of this population was heralded by many
as the discovery of the major epoch of dust-enshrouded 
spheroid formation (Lilly et al. 1999, Dunlop 2001, Granato et
al. 2001). This interpretation is by no means a consensus,
however. There have also been suggestions that much of the sub-mm
emission may originate from an extended cirrus component at lower
redshift (Rowan-Robinson 2000).

On a similar timescale, our understanding of the link between AGN and
massive galaxies has also taken a leap forward. Thanks largely to the
Hubble Space Telescope, it has become clear that essentially every
massive galaxy in the local Universe hosts a supermassive black hole
(Kormendy \& Richstone 1995, Magorrian et al. 1998). In particular,
the tight relationship between the black hole mass and the spheroidal
velocity dispersion suggests a possible link between an early epoch of
quasar activity and the formation of the spheroid (Gebhardt et
al. 2000, Ferrarese \& Merritt 2000).  So do the two phases coincide,
or is there a evolutionary sequence from one to the other, as appears
to be the case for local ULIRGs (Sanders et al. 1988)?  This is a
major unanswered question. One possibility is that the peak AGN
activity corresponds to a termination of the major burst of
star-formation activity producing the host spheroid (Silk \& Rees
1998, Fabian 1999). It has also been pointed out that a black hole
requires a significant length of time to grow to a sufficient size to
power a quasar, which may naturally lead to a lag between the epoch of
peak star formation and a subsequent quasar phase (Archibald et
al. 2001b).

There are other reasons for expecting a strong link between AGN and
SCUBA sources.  Models for the X-ray background require the existence
of a vast population of heavily absorbed AGN (Comastri et
al. 1995). There are major uncertainties in the models (in particular
the redshift and luminosity distribution of the obscured population)
but if this absorbed energy is re-radiated in the far infrared one
would expect $10-20$ per cent of the SCUBA sources to contain AGN
(Almaini, Lawrence \& Boyle 1999, Fabian \& Iwasawa 1999, Gunn \&
Shanks 1999).  The analogy with local ULIRGs is also instructive. At
the luminosities of the SCUBA sources ($> 10^{12}L_{\odot}$, assuming
$z\simeq2$) approximately $50$ per cent of local ULIRGs show clear
evidence for AGN activity (Sanders \& Mirabel 1996). The relative
roles of AGN and starburst activity in heating the dust remain
controversial however (Genzel et al. 1998; cf. Vignati et al. 1999).

Results from the first SCUBA surveys have already given some
indication of the likely fraction which contain AGN. The first
SCUBA-selected source to be optically identified was found to be a
gas-rich QSO (Ivison et al.\ 1998; Frayer et al.\ 1998).  Identifying
subsequent sub-mm sources has been exceptionally difficult, but
optical spectroscopy and SED constraints suggest that a sizeable
fraction ($\sim 30$ per cent) of the bright sub-mm sources are hosts
to AGN activity (e.g.\ Ivison et al. 2000, Cooray 1999). In contrast,
however, the first joint Chandra/SCUBA observations found a very small
overlap between the X-ray and sub-mm populations (Fabian et al.\ 2000;
Severgnini et al.\ 2000; Hornschemeier et al. 2000; cf.\ Bautz et al.\
2000), although there are some indications that a higher AGN fraction
is detectable in the deepest X-ray observations (Alexander et
al. 2001). In this paper we explore the X-ray/sub-mm connection using
Chandra observations of the largest contiguous sub-mm survey to
date. Being somewhat brighter than previous sub-mm surveys this should
allow tight constraints on the likely AGN content. The contiguous
nature of this survey is also an ideal match with the Chandra field of
view, allowing us to investigate the clustering between these
populations for the first time.

\begin{figure*} 
\centerline{\psfig{figure=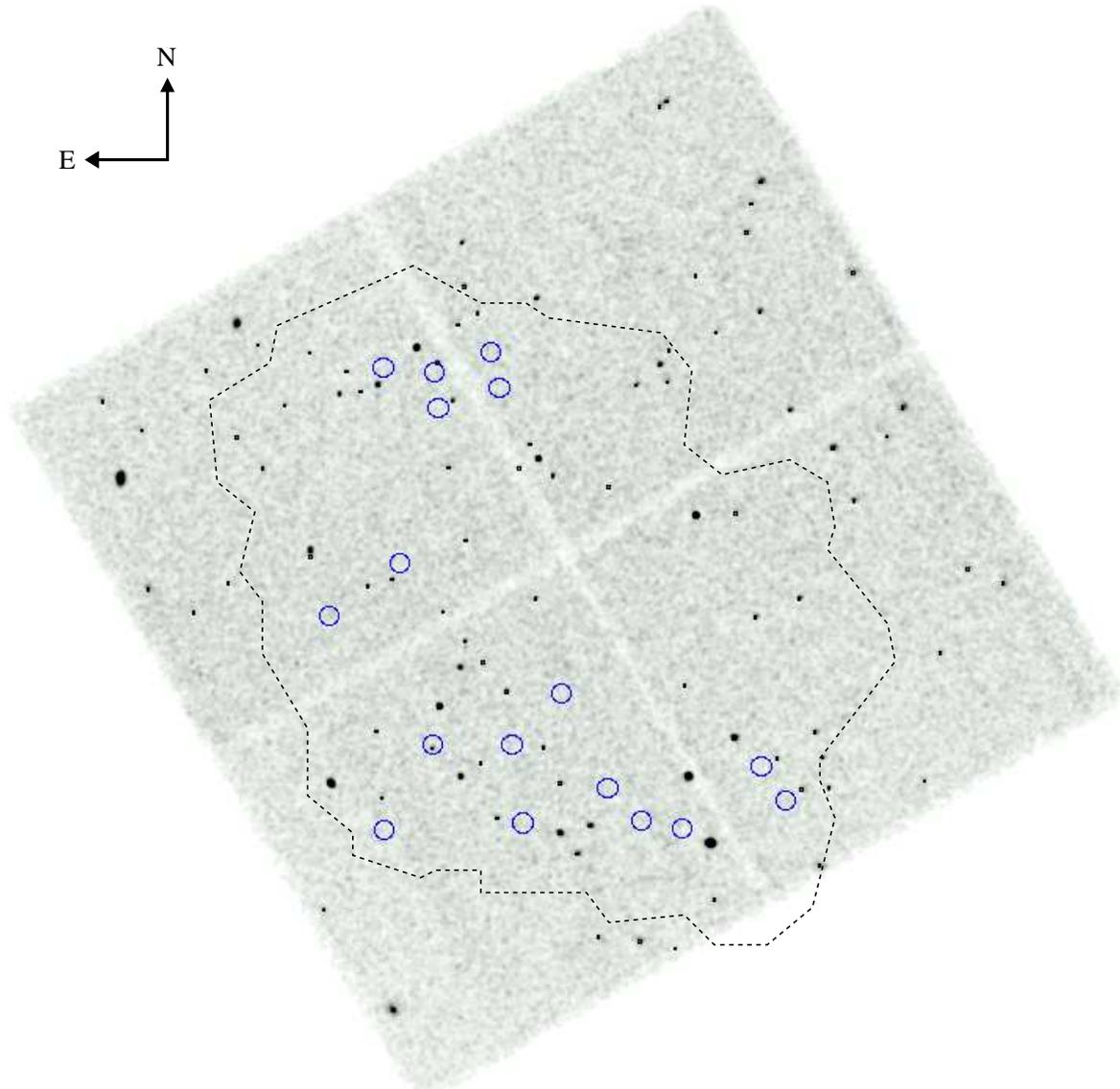,width=0.95\textwidth,angle=0}}
\caption{ $16.9\times16.9$ arcmin Chandra image (75~ks exposure) with
the positions of 17 SCUBA sources ($>3.5\sigma$) overlaid from the
8mJy survey (with large, $10$ arcsec radius error circles for
illustrative purposes). The dashed line indicates the extent of the
SCUBA coverage. The Chandra field is centred on ${\rm 16^h36^m47.0^s
+41^{\circ}01'34''}$ (J2000).  }
\end{figure*}

\begin{figure*}
\centerline{\psfig{figure=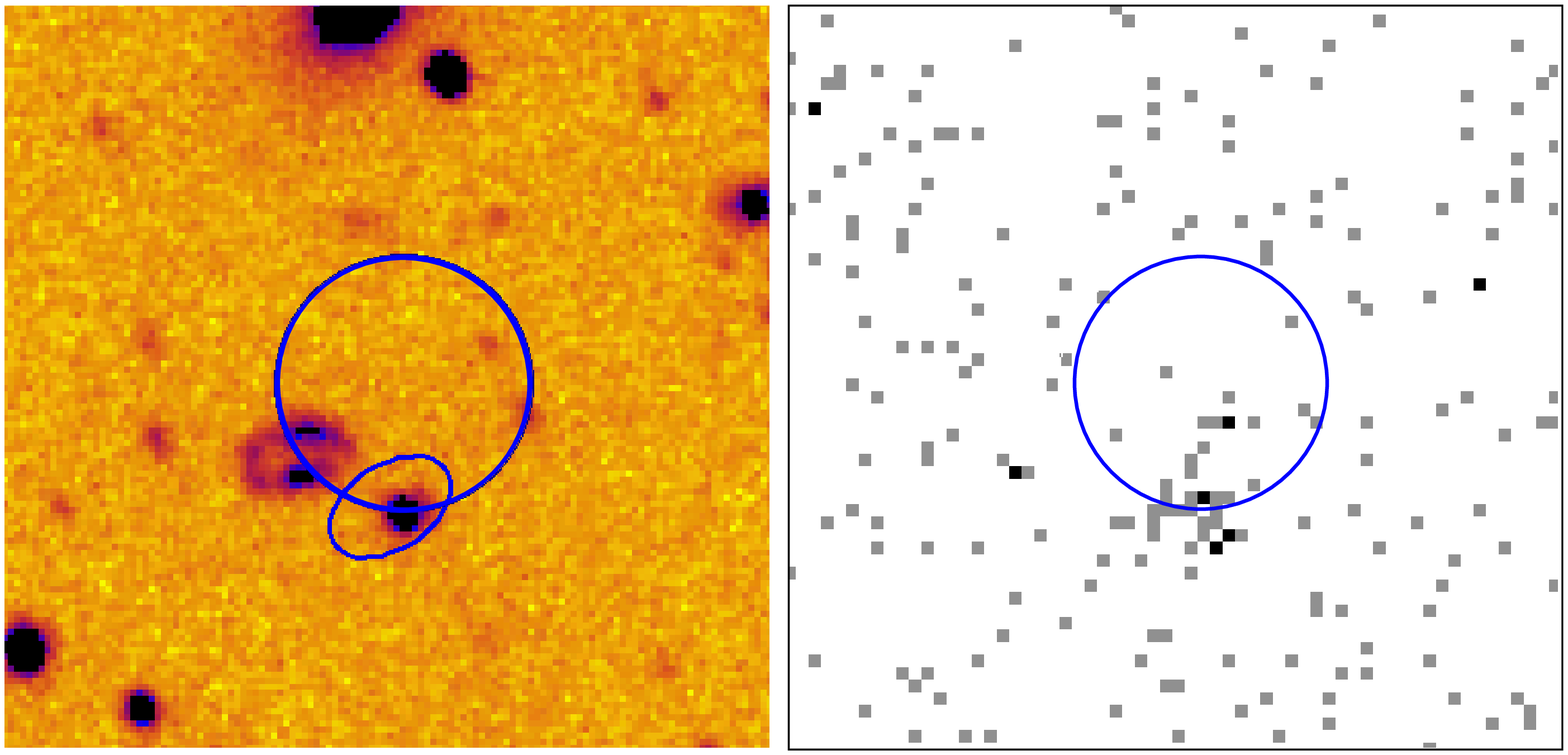,width=0.9\textwidth,angle=0}}
\caption{Optical R-band image of the region surrounding the SCUBA
source $N2\_850.8$ (left) with a $5$-arcsec radius SCUBA error
circle. This image has a magnitude limit of $R\simeq 26$. On the edge
of this error circle lies the faint Chandra source N2:19 (ellipse)
which has an optical counterpart with $R=22.45$.  It is unclear
whether this, the neighbouring merging system or a fainter
unidentified system is the true SCUBA source, although VLA
observations favour the Chandra position. The unsmoothed Chandra image
is shown on the right with $0.5$ arcsec pixels.  }
\end{figure*}

\section{The Observational Data}

\subsection{The SCUBA Survey}

The 8mJy SCUBA survey is the largest extragalactic sub-mm survey
undertaken so far, covering $260$ arcmin$^2$ in two regions of sky
(ELAIS N2 and the Lockman Hole East) to a typical rms noise level of
$\sigma=2.5$~mJy at $850~\mu$m.  Full details of the sub-mm observations
can be found in Scott et al. (2002). Details of the subsequent
multiwavelength follow-up can be found in Fox et al. (2002), Lutz et
al. (2001) and Ivison et al. (2002).  A significant
advantage over previous surveys is the source extraction method, based
on a data reduction pipeline which produces maps with independent
pixels and uncorrelated noise images (Serjeant et al. 2002). Using a
maximum likelihood technique, these enable the statistical
significance of each peak in the image to be assessed.  Combined with
a series of Monte-Carlo simulated maps one can also examine the
effects of completeness, confusion and the likely fraction of spurious
sources.

We adopt the $3.5\sigma$ SCUBA catalogue of Scott et al. (2002) as the
basis for this analysis. This corresponds to $17$ sources within the
Chandra region.  

\subsection{The ELAIS Deep X-ray Survey}

The N2 field is a particularly well studied region of sky, one of two
fields observed by Chandra from the European Large Area ISO Survey
(ELAIS, Oliver et al. 2000). Chandra observed this field on 2000
August 2 for 75~ks using the $2\times2$ array of ACIS-I CCDs.
Analysis of the X-ray data reveals $91$ sources within the $16.9\times
16.9$ arcmin field, to a flux limit of $5\times
10^{-16}~$erg$\,$s$^{-1}$cm$^{-2}$ ($0.5-8.0$~keV), of which $55$ lie
within the SCUBA map.  Approximately $90$ per cent of these Chandra
sources have an optical identification to a limit of $R\simeq26$. Our
deep optical imaging was also used to secure the X-ray astrometry to
an accuracy of $\simeq 0.5$ arcsec rms.  Full details of the X-ray
catalogue, source counts and hardness ratios can be found in Manners
et al. (2002). The optical/IR identification, photometry and
preliminary spectra can be found in Gonz\'alez-Solares et al. (2002)
and Willott et al. (2001).  This survey will soon be complemented by a
deep XMM observation ($150$ks). Combined with the sub-arcsec positions
of Chandra, this will allow us to push significantly deeper, in
addition to providing meaningful X-ray spectra and light curves.

The full-band Chandra image, smoothed with a 2-arcsec Gaussian, is
shown in Fig. 1. The positions of the $17$ SCUBA sources detected
above a threshold of $3.5\sigma$ are overlaid.

\section{The Chandra/SCUBA overlap}

Of the $17$ SCUBA sources detected above $3.5\sigma$, only one
($N2\_850.8$) contains a Chandra source within a $5$ arcsec error
radius (corresponding to a $90$ per cent positional error at
$850\mu$m). This is shown in Fig. 2.  Based on the density of Chandra
sources ($1200$deg$^{-2}$), we expect only $0.13$ of the $17$ SCUBA
error circles to contain a Chandra source by chance.  By Poisson
statistics, this corresponds to an $11$ per cent probability of at
least one chance alignment (although this could be increased if the
populations are correlated).

The SCUBA source $N2\_850.8$ has a flux of $5.1\pm1.4$~mJy at
$850~\mu$m.  The associated X-ray source has a flux of $4\pm1\times
10^{-15}~$erg$\,$s$^{-1}$cm$^{-2}$ ($0.5-8.0$~keV). It lies on the
edge of the SCUBA error circle, and coincides with a marginally
resolved compact optical counterpart at $R=22.45$.  The ring galaxy
$3$ arcsec from the X-ray position is also a plausible SCUBA
identification (reminiscent of the system discovered by Soucail et
al. 1999). Given this, and the $\sim 10$ per cent probability of
chance alignment, we view this Chandra/SCUBA coincidence with some
caution. We note, however, that our deep VLA imaging of this field
reveals a counterpart which coincides very well with the Chandra
position, with a flux density of $108\pm40~\mu$Jy at $1.4$~GHz (Ivison
et al., in preparation).  A photometric redshift based on $griHK$
photometry suggests a redshift of $z\simeq1.1$, with $90$ per cent
confidence limits in the range $0.85<z<1.15$ (assuming either a
heavily reddened young starburst or an elliptical galaxy SED). The
X-ray source is relatively hard, although fairly typical for a source
at this faint flux (see Manners et al. 2002 for a comparison of
hardness ratios).  Assuming $z=1.1$, the X-ray spectrum is consistent
with an absorbing column of $5\pm2 \times 10^{22}$cm$^{-2}$ (for an
intrinsic power law with $\alpha=0.7$). Optical spectroscopy combined
with sub-mm/mm interferometry may ultimately reveal the true
counterpart to this SCUBA source.

For the remaining SCUBA sources we estimate $3\sigma$ X-ray upper
limits of $4-7\times 10^{-16}~$erg$\,$s$^{-1}$cm$^{-2}$
($0.5-8$~keV). Further details of the Chandra flux limit and its
spatial dependence can be found in Manners et al. (2002). Accurate
SCUBA positions lead to tighter constraints on the X-ray flux in some
cases, particularly where our 1.4GHz VLA radio observations coincide
with a plausible optical/IR identification (Ivison et al. 2002).

\subsection{X-ray stacking analysis}

Although only $1/17$ SCUBA sources are detected by Chandra, we
performed a stacking analysis on the remaining $16$ sources to
estimate their mean X-ray flux. Extracting X-ray data using a
$7$-arcsec radius around each sub-mm source (corresponding to a
$2\sigma$ SCUBA error circle) we produced a Chandra image with an
effective exposure time of $1.2$Ms. This gave a weak detection of
X-ray flux at the level $1.1\pm0.4\times
10^{-15}~$erg$\,$s$^{-1}$cm$^{-2}$ ($0.5-8.0$~keV), corresponding to a
mean flux of $6.9\pm2.5\times 10^{-17}~$erg$\,$s$^{-1}$cm$^{-2}$
($0.5-8.0$~keV) per SCUBA source.

\section{Comparison with template SEDs}

\begin{figure*} 
\centerline{\psfig{figure=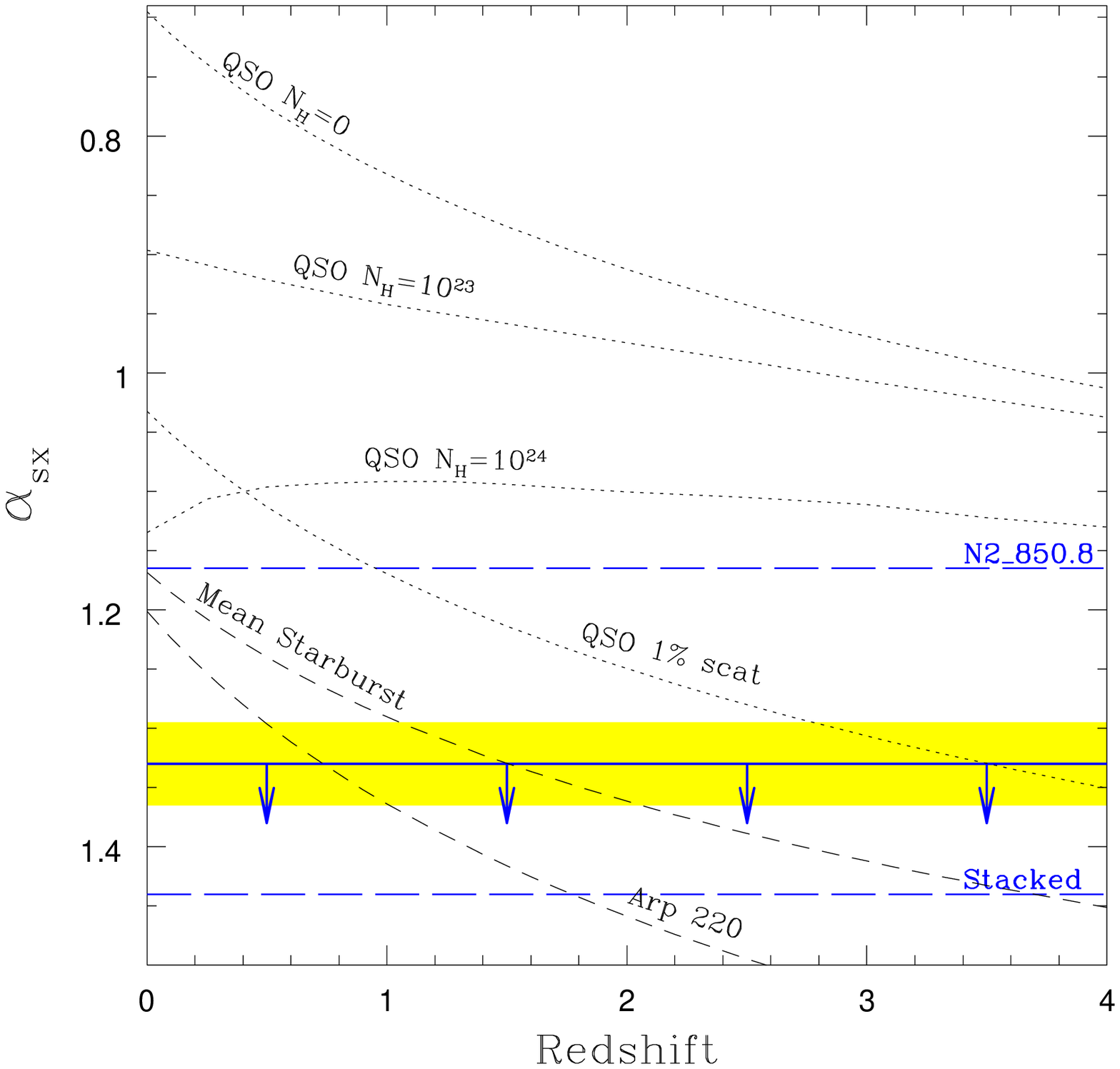,width=0.8\textwidth,angle=0}}
\caption{The observed sub-mm to X-ray spectral index ($\alpha_{SX}$)
as a function of redshift. The dotted curve shows the predictions for
radio-quiet quasars with range of photoelectric absorbing columns,
plus a simple Compton thick model in which only $1$ per cent of the
nuclear flux emerges.  The dashed curves show the predicted tracks for
a typical starburst galaxy, based on a the mean SED compiled by
Schmitt et al. (1997), and (for comparison) an X-ray-quiet starburst
such as Arp-220 . Further details of these models can be found in
Section 4.  The horizontal lines represent the spectral indices for
the $1$ detected SCUBA source ($N2\_850.8$) and the results of a
stacking analysis for the remainder.  Limits for individual SCUBA
sources are denoted by the shaded region.}
\end{figure*}

As described in Section 3, only one of the $17$ SCUBA sources
coincides with a Chandra source. The vast majority do not yield any
X-ray emission, although a stacking analysis yields a measurable X-ray
flux.  To convert these limits and detections into quantitative
statements on the AGN activity present we compare these limits with
the observed SEDs of local AGN and starburst galaxies. Fig. 3 displays
the expected sub-mm ($850~\mu$m) to X-ray ($2$~keV) spectral index as
a function of redshift for a number of template SEDs. The derivation
of these SEDs is described below.

\subsection{The quasar template}

For the quasar SED, we start with the $60~\mu$m to 2~keV spectral index
for radio-quiet quasars, based on the multiwavelength observations of
the Palomar-Green (PG) Bright Quasar Survey (Sanders et al. 1989).
Longward of $60~\mu$m the far-infrared/submillimetre emission is almost
certainly dominated by thermal re-radiation from dust (Carleton et
al. 1987, Hughes et al. 1993).  To model this spectral shape, we use
the best fitting parameters obtained by Hughes, Davies \& Ward (in
preparation), based on the largest collection of far-infrared and
submillimetre observations of local AGN.  They find that the
$50-1300\mu$m emission is consistent with thermal re-radiation from
dust at a temperature of $35-40$~K. This can be modeled by an
isothermal grey-body curve:

\begin{equation}
f_\nu \propto \frac{\nu^{3+\beta}}{\exp(h\nu/kT) -1} \hspace{1.2cm}
\lambda > 60 ~\mu m
\end{equation}

in which $T$ is the temperature of the dust and $\beta$ is the
emissivity index, derived by assuming that the grey-body is
transparent to its own emission.  Best fitting values were found to be
$\beta=1.7\pm0.3$ and $T=37\pm5$~K. 

In the X-ray regime, for an unabsorbed AGN we adopt a spectral index
with $\alpha=1.0$, consistent with X-ray studies of broad-line AGN
(e.g. Almaini et al. 1996, Reeves et al. 1997). To model highly
absorbed AGN, we simulate the effect of photoelectric absorption using
the XSPEC spectral analysis package, with large neutral hydrogen
columns up to $10^{24}$atom cm$^{-2}$. Much beyond this limit the
material will become Compton thick. Predicting the exact flux at
$2$~keV is somewhat meaningless given the faintness of the Chandra
sources, so we fold the output spectra through the full Chandra
response to predict the broad-band $0.5-8\,$~keV flux.  The model
values of $\alpha_{SX}$ are then calculated by converting the
full-band flux back to 2~keV, assuming a mean X-ray spectral index of
$\alpha=0.7$ (as assumed for any detected sources).

Fig. 3 illustrates the resulting sub-mm to X-ray predictions as a
function of redshift. These are calculated for a naked quasar
($N_H=0$), AGN with large absorbing columns ($N_H=10^{23}, 10^{24}$)
and finally a simple Compton-thick model in which only $1$ per cent of
the nuclear emission emerges through scattering. It is interesting to
note that for a very large absorbing column ($N_H\sim10^{24}$) the
negative K-corrections in both the X-ray and sub-mm wavebands
effectively cancel out, leaving a flux ratio essentially unchanged
with redshift.

\subsection{The starburst template}

Starburst galaxies are also luminous sources of hard X-ray emission,
although typically $2-3$ orders of magnitude lower than AGN.  The
primary source of hard X-ray emission is from short-lived massive
X-ray binaries and supernovae (Helfand et al. 2001, Natarajan \&
Almaini 2000). Studies of local starbursts indicate a tight
relationship between the far-infrared and X-ray luminosities (David,
Jones \& Forman 1992), suggesting that the X-ray luminosity is a
strong tracer of the underlying star formation.  To predict the
observed sub-mm to X-ray ratio, we use the mean $60\mu$m to X-ray
ratio from a recent compilation of starburst SEDs obtained by Schmitt
et al. (1997). We note that previous authors have used the spectrum of
Arp220 as a template (Fabian et al. 2000, Severgnini et al. 2000)
which leads to significantly lower X-ray predictions.  However it
should be stressed that Arp220 is anomalously weak in X-ray emission
compared to other luminous infrared galaxies (Iwasawa et al. 2001) so
we do not consider it to be representative. Nevertheless, we include 
predictions based on the the SED of this galaxy for comparison.

In the X-ray regime we assume a typical spectral index with
$\alpha=0.7$, consistent with the spectrum of well-studied local
starbursts (Moran, Lehnert \& Helfand 1999).  In the
far-infrared/sub-mm we scale from the $60\mu$m emission assuming the
same grey-body spectral shape as the AGN.  As noted by Lawrence
(2001), beyond $60\mu$m the far-infrared/sub-mm spectra of AGN and
starburst galaxies are essentially identical.  The values of $\beta$
and $T$ that we adopt are also in good agreement with SCUBA
observations of local infrared galaxies (Dunne et al. 2000).  The
resulting sub-mm to X-ray spectral indices as a function of redshift
are illustrated in Fig. 3.

\subsection{A comparison}

The only SCUBA source detected by Chandra ($N2\_850.8$) yields a value
of $\alpha_{SX}$ which is consistent with a highly absorbed AGN. An
unabsorbed QSO spectrum is strongly ruled out.  It is, however,
significantly more X-ray luminous than expected for a starburst galaxy
at $z>1$.

For the remaining SCUBA sources, an inspection of Fig. 3 yields a
number of interesting conclusions:

\newcounter{count2}
\begin{list}
{(\roman{count2})}{\usecounter{count2}}
\item{The SCUBA sources not detected by Chandra cannot be powered by
AGN unless the X-ray emission is completely absorbed by Compton thick
material. For redshifts $z<3$ any emerging scattered fraction must be
lower than $1$ per cent.}

\item{All of the SCUBA sources not detected by Chandra are consistent
with the predicted X-ray properties of starburst galaxies at $z>2$.
The limits on $\alpha_{SX}$ are in fact so strong that one can
plausibly use their X-ray non-detection as a redshift constraint. A
starburst of this sub-mm flux at $z<1$ should have been detected by
Chandra. The observed local scatter on the X-ray to FIR properties of
starbursts is not insignificant (David, Jones \& Forman 1992), but the
non-detection of $>90$\% of SCUBA sources suggests (at least
statistically) that the majority must lie at $z>1$. We note, however,
that a cooler cirrus contribution to the sub-mm luminosity could
weaken these redshift constraints (Rowan-Robinson 2000), as would the
prevalence of anomalously X-ray-weak objects like Arp220.  }
\end{list}

\begin{figure} 
\centerline{\psfig{figure=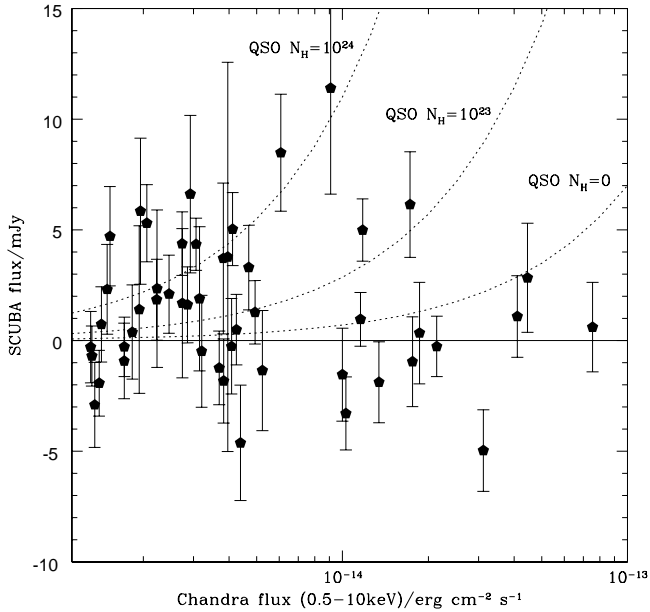,width=0.55\textwidth,angle=0}}
\caption{Estimates of the $850\mu$m flux for the $55$ Chandra sources
lying within the SCUBA map, displayed as function of X-ray flux. The
mean flux is $1.25\pm0.4$mJy. For comparison, models are shown based
on a quasar SED at $z=1.5$ with a range of absorbing columns. }
\end{figure}

\section{SCUBA limits on the  Chandra sources}

In addition to obtaining X-ray limits on the SCUBA sources, one can
reverse the argument and use this large SCUBA survey to obtain sub-mm
limits on the $55$ Chandra sources that lie within the SCUBA map
region. Based on typical AGN SEDs, we anticipate that most of the
Chandra sources will be too faint for detection by SCUBA in this
relatively shallow survey, but for completeness we estimate SCUBA
fluxes at the position of each Chandra source.  These were obtained by
$\chi^2$ fitting, using the source extraction algorithm discussed in
Scott et al. (2001). The results are displayed in Fig. 4.  Only the
Chandra source discussed in Section 3 yields a significant SCUBA
detection ($>3.5\sigma$), although others are detected at the
$2-3\sigma$ level.  The majority yield $3\sigma$ upper limits of
$<8$mJy.  Based on our current data, these sub-mm properties are
broadly consistent with predictions based on absorbed quasar SEDs (see
Fig. 4).  Redshifts and deeper sub-mm photometry will allow us to
investigate these SEDs further.

By co-adding the SCUBA beams at the positions of each Chandra source,
we obtain a mean sub-mm flux of $1.25\pm0.4$mJy.  A noise-weighted
mean gives $0.89\pm0.3$mJy.  To test the significance of these
results, the same analysis was performed after shifting the Chandra
sources to random positions within the SCUBA map. This was repeated
$100$ times. Reassuringly, a null result is obtained in most cases,
but the distribution of mean values suggests a $1-2$ per cent
probability of producing a `$3\sigma$' detection by chance. Deeper
sub-mm observations are clearly required to obtain significant
detections of individual Chandra sources, but we note that these
stacked results are consistent with the recent work of Barger et
al. (2001b).

\section{The clustering of SCUBA and Chandra sources}
\begin{figure} 
\centerline{\psfig{figure=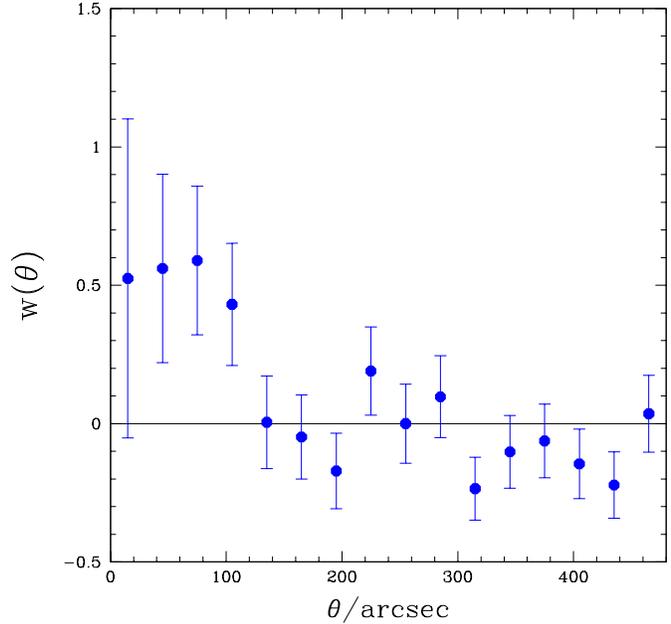,width=0.55\textwidth,angle=0}}
\caption{Statistical cross-correlation of the 17 SCUBA sources and the
complete sample of Chandra sources (see text). Within $100''$ we
obtain $82$ pairs compared to $51$ expected from a random
distribution. This represents a $4.3\sigma$ excess.  }
\end{figure}

\subsection{The cross-correlation}

Although the direct overlap between the SCUBA and Chandra sources is
small, an apparent clustering signal is visible by
eye (see Fig. 1). Much of this may be due to the `hole' in the
centre of the field, which is visible in both the Chandra and SCUBA
data.

To quantify this clustering signal we use a two-point
cross-correlation function $w_{SX}(\theta)$ of sub-mm and X-ray
sources.  This was obtained by counting the number of X-ray sources in
successive annuli  around the SCUBA sources. The
resulting radial distribution was then compared with the counts
obtained by placing $100,000$ random points over the Chandra field
area (avoiding edges and the chip boundaries).  $w_{SX}(\theta)$ can
then be obtained as follows:

\begin{equation}
w_{SX}(\theta_{i})=\frac{N_{SX}(\theta_{i})N_{R}}{N_{SR}(\theta_{i})N_{X}} -1
\end{equation}

where $N_R$ is the total number of random points, $N_X$ the number of
X-ray sources and {$N_{SX}(\theta_{i})$ and { $N_{SR}(\theta_{i})$
give the number of SCUBA/X-ray and SCUBA/random pairs respectively.

Unfortunately the sensitivity of Chandra degrades with radius,
particularly towards the corner of the chips (a factor of two
by $10$ arcmin). This alters the flux limit, which may lead to a false
gradient in the number density of Chandra sources. To mitigate these
effects, we restrict the analysis to the central $7$ arcmin radius
from the Chandra focus.

The resulting cross-correlation signal is displayed in Fig. 5, with
Poisson error bars. A positive signal is observed at small angular
separations. Within a radius of $100''$ we obtain $82$ pairs compared
to $51$ expected from a random distribution. This represents a
$4.3\sigma$ detection of clustering.

\begin{figure} 
\centerline{\psfig{figure=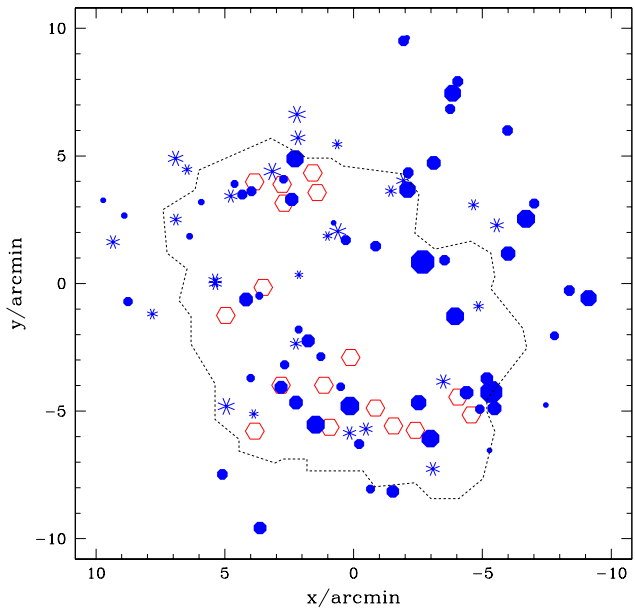,width=0.5\textwidth,angle=0}}
\caption{Showing the relative positions of the SCUBA sources (unfilled
hexagons) and Chandra identifications. The dashed line indicates the
extent of the sub-mm coverage.  Unlike Fig. 1, the size of the Chandra
points are proportional to the log of their optical (R-band)
flux. Filled points represent Chandra sources with resolved optical
identifications, while the stars illustrate Chandra sources with
stellar counterparts. The largest filled points are mostly low
redshift galaxies, while the smaller filled points and stars should be
predominantly high-$z$ AGN and quasars ($1<z<4$). }
\end{figure}

\begin{figure} 
\centerline{\psfig{figure=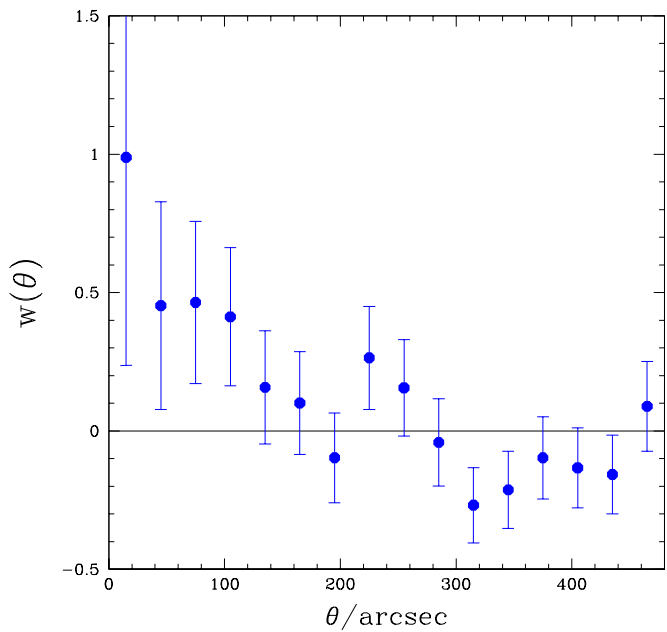,width=0.55\textwidth,angle=0}}
\caption{Statistical cross-correlation of the 17 SCUBA sources and the
`high-redshift' subset of Chandra sources (see text).}
\end{figure}

\subsection{The high-redshift Chandra sources}

A problem which plagues all sub-mm surveys is the extreme optical
faintness of the sources (Smail et al. 2002). Typically most are
invisible even in the deepest optical imaging. For this
reason, none of the SCUBA sources in our survey have a confirmed
redshift. Nevertheless, a combination of arguments lead to the
conclusion that the majority must lie at $z>1.5$. This is based on
their $450/850\mu$m ratios, deep VLA radio observations and their
optical/IR properties (Smail et al. 2000, Dunlop 2000, Fox et
al. 2002, Ivison et al. 2001). The Chandra sources, however, are known
to have a broad redshift distribution ($0<z<4$), including the newly
discovered population of weak AGN in low redshift spheroidal galaxies
(Mushotzky et al.  2000).

To investigate the clustering signal further we therefore repeat
Chandra/SCUBA cross-correlation with a `high-redshift' Chandra sample.
So far we have only $\sim 15$ redshifts for Chandra sources in this
field.  The majority lie at $z>1.5$, but we exclude the $2$ examples
which lie at $z<1$ (Gonzales-Solares et al. 2001). Further redshifts
will be obtained, but in the meantime we can exclude many obvious
low-$z$ AGN from our sample.  Formally, we exclude all Chandra sources
with bright optical IDs ($R<22$) which are also clearly resolved as
galaxies. Based on the redshift distribution of galaxies at this
magnitude (Cohen et al. 2000) and the identifications from the first
deep Chandra surveys (Hornschemeier et al. 2001, Barger et al. 2001a,
Tozzi et al. 2001) these should lie predominantly at $z<1$. In total
this excludes $19$ of the $91$ Chandra sources. Repeating the
cross-correlation with the remaining `high-$z$' sample does not
increase the amplitude of $w(\theta)$ significantly however (see
Fig. 7). Formally the significance actually drops to a $3.8\sigma$
excess within $100''$.

We conclude that while high-$z$ Chandra sources account for most of
the cross-correlation signal, at least part of this effect may be
produced by SCUBA sources correlated with structure at low redshift
($z<1$).  This can be seen visually in Fig. 6. Could some fraction of
the SCUBA sources lie at low redshift, or is this caused by
gravitational lensing? We investigate this effect further in a related
paper (Almaini et al., in preparation) by cross-correlating SCUBA
positions with optical galaxy catalogues.

\begin{figure*} 
\centerline{\psfig{figure=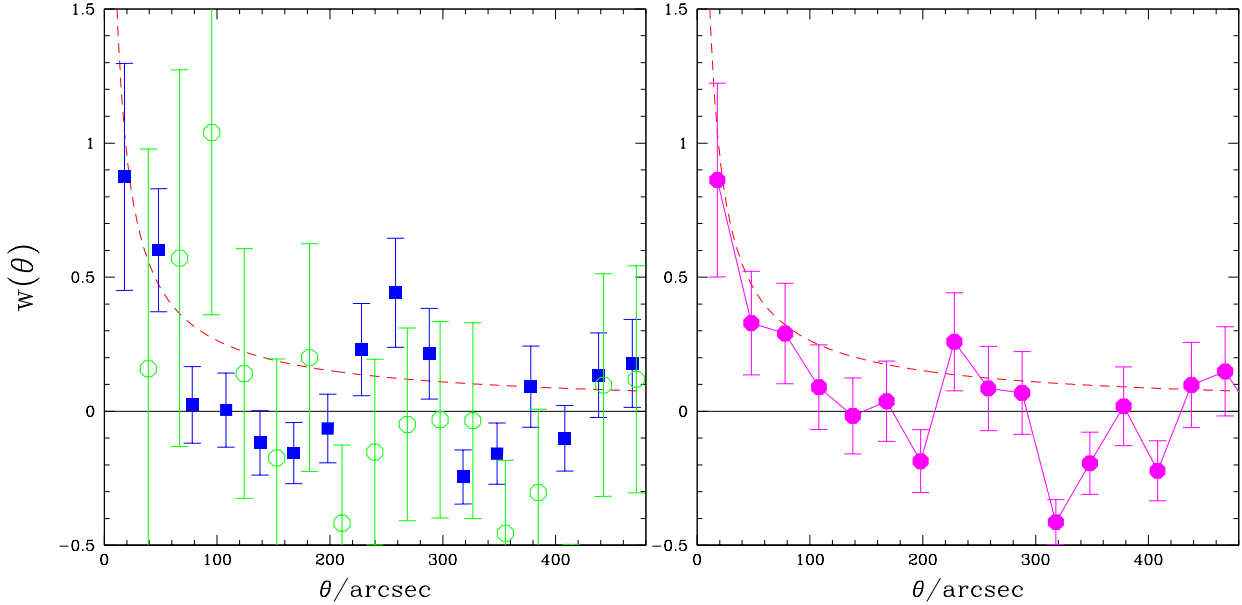,width=1.0\textwidth,angle=270}}
\caption{On the left we display the auto-correlation function for the
SCUBA sources (circles) and `high-redshift' Chandra sources
(squares). On the right the auto-correlation for the combined sample
is displayed. For comparison, the dashed lines shows the best-fitting
correlation function for EROs (Daddi et al. 2000).}
\end{figure*}

\subsection{The auto-correlation functions}

To investigate clustering within the two populations separately, we
evaluate a two-point {\em auto}-correlation function. This has the
same form as equation $(2)$ but is based on a self-correlation within a
given catalogue. The auto-correlation function for the SCUBA sources
was originally evaluated in Scott et al. (2001) and is reproduced
here. A comparison with the $w(\theta)$ for the `high-redshift'
Chandra sources is shown in Fig. 8(a).  The results are somewhat
tentative, with neither population revealing a conclusive detection of
clustering.

The strength of the {\em cross}-correlation signal, however, suggests
that a significant fraction of these populations are tracing the same
structure (Section 5.1). We therefore assume that these are all
`massive galaxies' of one form or another, combine the `high-z'
Chandra sources with the SCUBA sources, and evaluate the
auto-correlation function for the joint sample. Not surprisingly, the
resulting angular auto-correlation function is noisy, but we obtain a
$3.8\sigma$ detection of clustering within $100$ arcsec (Fig. 8b). The
best fitting function, adopting the standard functional form
$w(\theta) = A \theta^{-0.8}$, gives an amplitude $A(1^{\rm
o})=0.011\pm0.004$.  This is strikingly similar to the amplitude of
clustering for bright ($K<18$) EROs (Daddi et al. 2000).  Given that
EROs are now widely believed to be dominated by passively evolving
elliptical galaxies at $z=1-2$, if the strength of the Chandra/SCUBA
clustering can be confirmed it may argue for a link between these
apparently disparate populations.  Could sub-mm sources, quasars and
elliptical galaxies represent massive galaxies at different stages in
their evolution?  Based on the small area, the small number of sources
and the tentative nature of this clustering signal we urge caution in
jumping to premature conclusions. Furthermore, the Chandra/SCUBA
population are likely to trace higher redshifts than the EROs from
Daddi et al. (2000). Different $N(z)$ distributions could lead to them
having different inferred degrees of bias, even if the angular
clustering is identical.  The strength of the clustering signal may
also be boosted by gravitational lensing (Almaini et al., in
preparation). Nevertheless, we believe that these results provide
motivation for a sub-mm/X-ray survey covering a significantly larger
area. The aim will be to exploit a very basic property of hierarchical
theories for structure formation, namely the strong dependence of
clustering on the halo mass (Baugh et al. 1998, Steidel et al. 1999,
Magliocchetti et al. 2001).

\section{Implications for the joint formation of spheroids and AGN}

The major new result of this work is the detection of a strong
cross-correlation between the Chandra and SCUBA sources. This
immediately suggests that these phenomena are tracing the same large
scale structure.  We have also confirmed the findings of other recent
authors, namely that the direct overlap between the bright SCUBA and
Chandra populations is small (at most $10$ per cent). At first sight,
these results appear somewhat contradictory.  If the SCUBA sources are
indeed the progenitors of massive elliptical galaxies then they should
eventually harbour exceptionally massive black holes (Kormendy \&
Richstone 1995, Magorrian et al. 1998). In particular, the tight
relationship between black hole mass and the mass of the spheroid
(Gebhardt et al. 2000, Ferrarese \& Merritt 2000) suggests a direct
link between the epoch of spheroid formation (the SCUBA phase?)  and
the growth of the black hole.  Our observations, however, suggest that
the two phenomena only coincide $\sim 5$ per cent of the time. There
are several plausible explanations for this low overlap:

\newcounter{count}
\begin{list}
{(\roman{count})}{\usecounter{count}}
\item{{\bf The nucleus is not receiving fuel.} 

With star-formation rates of $\sim 1000 M_{\odot}$yr$^{-1}$ (as
required to power the most luminous SCUBA sources) the resulting
stellar winds and supernovae should certainly generate a plentiful
supply of fuel. The transfer of this material to the sub-parsec scale
of an accretion disk, however, is by no means guaranteed (Shlosman,
Begelman \& Frank 1990). It therefore seems plausible that some
fraction of the SCUBA sources are simply not fueling the central black
hole at the observed epoch, although this would appear inconsistent
with the simplest explanation of the black-hole/spheroid mass
relationship (i.e. a joint growth process).  We also note however that
at least $50$ per cent  of local ULIRGs with equivalent luminosities show
clear evidence for an active nucleus (Sanders \& Mirabel 1996).}

\item{{\bf The quasar terminates the star-formation} 

A number of models have been proposed in which the onset of QSO
activity terminates the formation of the host spheroid (Silk \& Rees
1998, Fabian 1999, Granato et al. 2001). This would lead to a natural
time-lag between the two phenomena. }

\item{{\bf The black hole is growing.}

Archibald et al. (2001b) have proposed a new model in which the
central quasar is alive but still growing.  This is based on the
hypothesis that a black hole is likely to grow from a small seed
($\sim 100M_{\odot}$) and hence, even accreting at the Eddington
limit, it will require $\sim 5\times10^{8}$ years to reach a
sufficient size to power a quasar. At this stage the peak
star-formation activity may have ended, leading to a natural lag
between the SCUBA phase and the subsequent luminous quasar. An AGN
with $<1$ per cent of the final quasar luminosity would remain
undetected by Chandra, corresponding to Eddington limited accretion
onto a black hole of $\sim 10^7 M_{\odot}$ for a typical SCUBA source
at $z=3$. }

\item{{\bf The Compton-thick scenario.} 

If the obscuring column is $\gg 10^{24}$ atom cm$^{-2}$ the continuum
will be completely suppressed. Only nuclear emission scattered into
our line of sight will be observed, and if this is less than $\sim 1$
per cent it will remain invisible even at the depth of our Chandra
observations (see Fig. 3). Very low scattered fractions are not
uncommon in local AGN-dominated ULIRGs (Fabian et al. 1996).  It
should be noted, however, that a Compton-thick explanation cannot
occur within the context of a `Unified Scheme'  as applied to
local AGN. If all the SCUBA sources contain enshrouded AGN, the
detection of only $\sim 5$\% by Chandra would require very small
opening angles, i.e.  almost isotropic obscuration.  This is
consistent with models in which a nuclear starburst both fuels and
obscures the active nucleus (Fabian et al. 1998).  }
\end{list}

Clearly not all of these models are not mutually exclusive, e.g. the
AGN may be obscured by Compton thick material while the black hole is
growing.

\section{Summary and conclusions}

We present deep Chandra observations of a wide-area SCUBA sub-mm
survey.  In agreement with other recent findings, we confirm that the
direct overlap between the X-ray and bright sub-mm populations is
small. Only $1/17$ SCUBA sources are coincident with Chandra
detections. Coadding the X-ray flux for the remaining $16$ sources we
obtain a weak detection, but this is entirely consistent with
starburst activity.

By a detailed comparison with the SEDs of quasars and starburst
galaxies, we conclude that the majority of the bright SCUBA sources
are not powered by AGN, unless the central engine is obscured by
Compton-thick material with a low ( $<1$ per cent) scattered
fraction. It should be noted, however, that a Compton-thick
explanation cannot occur within the context of the `Unified Scheme'
scenario as applied to local Seyferts. If most SCUBA sources contain
enshrouded AGN, the detection of only $5$\% by Chandra would require a
very small typical opening angle, i.e.  almost isotropic obscuration.

The X-ray limits are so strong in most cases that we can also rule out
a starburst SED at low redshift. The non-detection of $16/17$ SCUBA
sources suggests (at least statistically) that the majority lie at
$z>1$ even if they are purely starburst galaxies. 

Reversing the perspective, only one of the $55$ Chandra sources lying
within the overlap region is significantly detected by SCUBA, although
several yield detections at the $2-3\sigma$ level.  The remainder show
typical $3\sigma$ upper limits of $<8$~mJy at $850~\mu$m, broadly
consistent with the expectations of absorbed AGN models. By co-adding
the SCUBA beams at the positions of each Chandra source, we obtain a
mean sub-mm flux of $1.25\pm0.4$mJy, consistent with Barger et
al. (2001b).

Despite the absence of X-ray emission from $95$ per cent of bright
sub-mm sources, we find evidence for strong angular clustering between
the Chandra and SCUBA populations ($4.3\sigma$ significance).  The
strength of this signal is consistent with the clustering seen among
EROs (Daddi et al. 2000).  The implication is that luminous AGN and
SCUBA sources trace the same large scale structure, but {\em for a
given massive galaxy} the quasar phase and the peak episode of star
formation do not coincide. This may be due to very heavy
(Compton-thick) obscuration of the central nucleus, but could also
reflect an evolutionary time-lag between the formation of the spheroid
and the onset of (visible) quasar activity.

\section*{ACKNOWLEDGMENTS}

We thank Ian Smail, Richard Ellis and Scott Chapman for useful
discussions. We also wish to thank an anonymous referee for some very
useful suggestions. OA acknowledges the considerable support offered
by the award of a Royal Society Research Fellowship. JSD acknowledges
the enhanced research time afforded by the award of a PPARC Senior
Fellowship.


\begin{thebibliography}{}

\bibitem[]{}
Alexander D. et al.. 2001, ApJ, 122, 2156

\bibitem[]{}
Almaini O., Lawrence A. \& Boyle B.J., (1999), MNRAS, 305, L59

\bibitem[]{}
Almaini O. et al.,   1996, MNRAS, 282, 295

\bibitem[]{} 
Archibald E.N., Dunlop J.S., Hughes D.H., Rawlings S.,
Eales S.A., Ivison R.J., 2001a, MNRAS, 323, 417


\bibitem[]{} Archibald E.N., Dunlop J.S., Jimenez R., Friaca A.C.S.,
Mclure R.J., Hughes D.H., 2001b, MNRAS, submitted, 
astro-ph/0108122

\bibitem[]{} Barger A.J., Cowie L.L., Sanders, D.B., Fulton E.,
Taniguchi Y., Sato Y., Kawara K,, Okuda H,, Nature, 394, 248



\bibitem[]{}
Barger A.J., Cowie L.L., Mushotzky R.F., Richards E.A., 
et al., 2001a, AJ, 121, 662

\bibitem[]{}
Barger A.J., Cowie L.L., Steffen A.T.,
Hornschemeier A.E., Brandt W.N., Garmie G.P., 
2001b, ApJ, 2001, ApJ, 560, L23

\bibitem[]{} 
Baugh C.M., Cole S., Frenk C.S., Lacey C.G., 
1998, ApJ, 498, 504

\bibitem[]{} 
Bautz M.W., Malm, M. R., Baganoff, F. K., Ricker, G. R.,
Canizares, C. R., Brandt, W. N., Hornschemeier, A. E., Garmire, G. P.,
2000, ApJ, 543, 119


\bibitem[]{}
Blain A.W., Longair M.S., 1993, MNRAS, 265, L21

\bibitem[]{}
Carleton N.P., Elvis M., Fabbiano G., Willner S.P.,
Lawrence A., Ward M., 1987, ApJ, 318, 595




\bibitem[]{}
Cohen J.G. et al., 2000, ApJ, 538, 29


\bibitem[]{}
Comastri A., Setti G., Zamorani G. \& Hasinger G.,
1995,  A\&A, 296, 1

\bibitem[]{}
Cooray A.R., 1999, New Astronomy, 4, 377



\bibitem[]{} 
Daddi, E., Cimatti, A., Pozzetti, L., Hoekstra, H.,
Röttgering, H.J.A., Renzini, A., Zamorani, G., Mannucci, F., 2000,
A\&A, 361, 535

\bibitem[]{} 
David L.P., Jones C. \& Forman W., 1992, ApJ, 388, 82


\bibitem[]{} 
Dunlop, J.S., 2001, in: Deep Sub-millimetre Surveys, eds.
Lowenthal, J. \& Hughes, D.H., World Scientific, in
press. (astro-ph/0011077)

\bibitem[]{} Dunne L., Eales S., Edmunds M., Ivison E., Alexander P.,
Clements D.L., 2000, MNRAS, 315, 115

\bibitem[]{}
Eales S. et al, 1999, ApJ, 517, 148

\bibitem[]{}
Fabian A.C., Cutri R.M., Smith H.E., Crawford C.S., Brandt W.N., 
1996, MNRAS, 283, L95

\bibitem[]{}
Fabian A.C., Barcons X., Almaini O., Iwasawa K., 1998, MNRAS, 297, L11

\bibitem[]{}
Fabian A.C., Iwasawa K., 1999, MNRAS, 303, 34

\bibitem[]{}
Fabian A.C., 1999, MNRAS, 308, L39

\bibitem[]{}
Ferrarese L., Merritt D., 2000, ApJ, 539, 9

\bibitem[]{}
Fox. M.J., et al., 2002, MNRAS, 331, 839


\bibitem[]{}
Frayer D.T., Ivison R.J., Scoville N.Z., Yun M., 
Evans A.S., Smail I., Blain A.W., Kneib, J.-P., 1998, 
ApJ, 506, L7


\bibitem[]{}
Gebhardt K., et al., 2000, ApJ, 539, 13


\bibitem[]{}
Genzel R., et al., 1998, ApJ, 498, 479

\bibitem[]{}
Gonz\'alez-Solares et al., 2001, in preparation

\bibitem[]{}
Granato G.L., Silva L., Monaco P., Panuzzo P., Saliccu P., 
De Zotti G,, Danese L., 2001, MNRAS, 324, 757

\bibitem[]{}
Gunn K.F. \& Shanks T., 1999, MNRAS submitted (astro-ph/9909089)

\bibitem[]{}
Helfand D.J. \& Moran E.C., 2001, ApJ, 554, 27

\bibitem[]{}
Hornschemeier A.E., et al., 2000, ApJ, 541, 49

\bibitem[]{}
Hornschemeier A.E., et al., 2001, ApJ, 554, 742

\bibitem[]{}
Hughes D.H., Robson E.I., Dunlop J.S., Gear W.K., 1993,
MNRAS, 263, 607

\bibitem[]{}
Hughes D.H. et al., 1998, Nature, 394, 241

\bibitem[]{} 
Ivison R.J., Smail I., Le Borgne J.F., Blain A.W., Kneib
J.P., Bezecourt J., Kerr T.H., Davies J.K., 1998, MNRAS, 298, 583

\bibitem[]{} 
Ivison R.J., Smail I., Barger A.J.,  Kneib
J.P., Blain A.W., Owen F.N., Kerr T.H., Cowie L.L., 2000, MNRAS, 315, 209


\bibitem[]{} 
Iwasawa K., Matt G., Guainazzi M., Fabian A.C., 2001,
MNRAS, 326, 894

\bibitem[]{}
Kormendy, J., \& Richstone, D. 1995, ARA\&A 33, 581

\bibitem[]{} 
Lawrence A., 2001, In proceedings from ESA Symposium,
``The Promise Of First'', Toledo 2000, Ed. G.L Pilbratt et al., 
(astro-ph/0105305)

\bibitem[]{}
Lilly S.J. et al. 1999, ApJ, 518, 641 

\bibitem[]{}
Lutz D., et al., 2001, A\&A, 378, 70

\bibitem[]{} Magliocchetti M., Moscardini L., de Zotti G., Granato
G.L., Danese L., 2001, to appear in "Where is the Matter? Tracing Dark
and Bright Matter with the New Generation of Large Scale Surveys",
June 2001, Treyer \& Tresse Eds, Frontier Group (astro-ph/0107597)

\bibitem[]{}
Magorrian, J., et al. 1998, AJ, 115, 2285 

\bibitem[]{}
Manners et al., 2002, MNRAS, submitted (astro-ph/0207622)


\bibitem[]{}
Moran E.C., Lehnert M.D. \& Helfand D.J., 1999, ApJ, 526, 649

\bibitem[]{}
Mushotzky R.F., Cowie L.L., Barger  A.J.,  Arnaud, K. A., 2000, 
Nature, 404, 459 

\bibitem[]{}
Natarajan P. \& Almaini O., 2001, MNRAS, 318, L21

\bibitem[]{}
Oliver S. et al., 2000, MNRAS, 316, 749

\bibitem[]{}
Reeves J.N., Turner M.J.L., Ohashi T., Kii T., 1997, MNRAS, 292, 468

\bibitem[]{}
Rowan-Robinsom, M., 2000, ApJ, 549,745



\bibitem[]{}
Sanders D.B., Soifer B.T., Elias J.H., Madore B.F., Matthews K., Neugebauer G., Scoville N.Z., 1988, ApJ, 325, 74


\bibitem[]{} 
Sanders D.B., Phinney E.S., Neugebauer G., Soifer B.T.,
Matthews K., 1989, ApJ, 347, 29

\bibitem[]{} Sanders D.B. \& Mirabel I.F., 1996, ARA\&A, 34, 749


\bibitem[]{}
Scott S.E., et al., 2002, MNRAS, 331, 817

\bibitem[]{}
Serjeant S. et al., 2002, MNRAS, n press (astro-ph/0201502) 

\bibitem[]{}
Severgnini P. et al., 2000, A\&A, 360, 457

\bibitem[]{}
Shlosman, I., Begelman, M.C.,  Frank, J., 1990, Nature, 345, 679

\bibitem[]{}
Silk. J \& Rees M.J., 1998, A\&A, 331, L1

\bibitem[]{}
Smail I., Ivison R.J., Blain A.W., 1997, ApJ, 490, L5


\bibitem[]{}
Smail I., Ivison R.J., Owen F.N., Blain A.W., Kneib J.P., 2000, ApJ, 528, 612

\bibitem[]{}
Smail I., Ivison R.J., Blain A.W. \& Kneib J.P., 2002, MNRAS, 331, 495

\bibitem[]{}
Soucail G., Kneib J.P., Bezecourt J., Metcalfe L., Altieri B., Le Borgne J.F., 1999, A\&A, 343, L70

\bibitem[]{}
Steidel, C.C., Adelberger, K.L., Giavalisco, M., Dickinson, M.,
Pettini, M., 1999, ApJ, 519, 1

\bibitem[]{}
Tozzi P. et al., 2001, ApJ, 562, 42

\bibitem[]{}
Vignati P., et al., 1999, A\&A 349, L57

\bibitem[]{}
Willott et al., 2001, MNRAS, submitted (astro-ph/0105560)

\end{thebibliography}
\end{document}